\documentclass[fleqn,usenatbib]{mnras}

\usepackage{newtxtext,newtxmath}
\usepackage{CJK}
\usepackage[T1]{fontenc}

\DeclareRobustCommand{\VAN}[3]{#2}
\let\VANthebibliography\thebibliography
\def\thebibliography{\DeclareRobustCommand{\VAN}[3]{##3}\VANthebibliography}

\usepackage{graphicx}	% Including figure files
\usepackage{amsmath}	% Advanced maths commands
\title[PHL~5038AB: a WD-BD binary system]{PHL~5038AB: Is the brown dwarf causing pollution of its white dwarf host star?}

\author[S. L. Casewell et al.]{
S. L. Casewell,$^{1}$\thanks{E-mail: slc25@leicester.ac.uk}
J. Debes,$^{2}$
T. J. Dupuy,$^{3}$
P. Dufour,$^{4}$
A. Bonsor,$^{5}$
A. Rebassa-Mansergas$^{6,7}$,\newauthor
R. Murillo-Ojeda$^{8}$,
J. R. French,$^{1}$
R. D. Alexander,$^{1}$
Siyi Xu \begin{CJK*}{UTF8}{gbsn}(许\CJKfamily{bsmi}偲\CJKfamily{gbsn}艺)\end{CJK*}$^{9}$,
E. Martin$^{10}$,
E. Manjavacas$^{2}$
\\
$^{1}$Centre for Exoplanet Research, School of Physics and Astronomy, University of Leicester, University Road, Leicester, LE1 7RH, United Kingdom\\
$^{2}$Aura for the European Space Agency (ESA), ESA Office, Space Telescope Science Institute, 3700 San Martin Drive, Baltimore, Maryland 21218, USA\\
$^{3}$Royal Observatory Edinburgh, Blackford Hill, Edinburgh, EH9 3HJ, United Kingdom\\
$^{4}$ De\'epartement de Physique, Universit\'e de Montr\'eal, C.P. 6128, Succ. Centre-Ville, Montr\'eal, Qu\'ebec H3C 3J7, Canada\\
$^{5}$Institute of Astronomy, University of Cambridge, Madingley Road, Cambridge, CB3 0HA, UK\\
$^{6}$ Departament de F\'isica, Universitat Polit\`ecnica de Catalunya, c/Esteve Terrades 5, 08860 Castelldefels, Spain\\
$^{7}$ Institute for Space Studies of Catalonia, c/Gran Capit\`a 2--4, Edif. Nexus 104, 08034 Barcelona, Spain\\
$^{8}$ Centro de Astrobiolog\'ia (CAB), CSIC-INTA, Camino Bajo del Castillo s/n, campus ESAC, 28692, Villanueva de la Ca\~nada, Madrid, Spain\\
$^{9}$Gemini Observatory/NSF's NOIRLab, 670 N. A'ohoku Place, Hilo, Hawaii, 96720, USA\\
$^{10}$Department of Astronomy \& Astrophysics, University of California Santa Cruz, 1156 High Street, Santa Cruz, California 95064, USA\\
}

\date{Accepted XXX. Received YYY; in original form ZZZ}

\pubyear{2024}

\begin{document}
\label{firstpage}
\pagerange{\pageref{firstpage}--\pageref{lastpage}}
\maketitle

\begin{abstract}
We present new results on PHL~5038AB, a widely separated binary system composed of a white dwarf and a brown dwarf, refining the white and brown dwarf parameters and determining the binary separation to be $66^{+12}_{-24}$~AU. New spectra of the white dwarf show calcium absorption lines suggesting the hydrogen-rich atmosphere is weakly polluted, inferring the presence of planetesimals in the system, which we determine are in an S-type orbit around the white dwarf in orbits closer than 17-32 AU. We do not detect any infrared excess that would indicate the presence of a disc, suggesting all dust present has either been totally accreted or is optically thin. In this system, we suggest the metal pollution in the white dwarf atmosphere can be directly attributed to the presence of the brown dwarf companion disrupting the orbits of planetesimals within the system.
\end{abstract}


%%%%%%%%%%%%%%%%%%%%%%%%%%%%%%%%%%%%%%%%%%%%%%%%%%

%%%%%%%%%%%%%%%%% BODY OF PAPER %%%%%%%%%%%%%%%%%%

\section{Introduction}
It is estimated that 25 to 50 per cent of white dwarfs have atmospheres polluted by metals \citep{Zuckerman03, zuckerman10, koester14, hollands17}. Due to the rapid sinking times of metals in these dense objects, in order for them to be detected, there must be a continual replenishing of the pollutants \citep{Koester09, barstow}. 

The favoured explanation for the origin of these pollutants is that planetesimals (similar to asteroids) orbiting the white dwarf on eccentric orbits are scattered inwards where they sublimate and eventually accrete onto the white dwarf (e.g. \citealt{debes02, Jura2003}). The composition of the disrupted objects is then inferred from the species present in the white dwarf atmosphere (e.g. \citealt{klein}). Further evidence for this scenario includes the discovery of planetesimals that transit the polluted white dwarf WD1145+017 \citep{vanderburg15}, and further discoveries around other white dwarfs (e.g \citealt{vanderbosch21}).   

A range of dynamical mechanisms has been suggested to transport these planetesimals to the inner regions of white dwarf planetary systems, including stellar mass loss via winds, planet(s), and wide-binary companions  (e.g. \citealt{Bonsor11, Debes12, Veras14,  Bonsor15, Portegies15, Petrovich2017, Hamers16, Mustill18}). However,  the most likely explanation is that they are scattered inwards by planets in a surviving outer planetary system, rather like sun-grazing comets in the Solar System.  Despite this suggestion, no system has yet been discovered where a definitive mechanism is identified as responsible.

While polluted white dwarfs are relatively common, white dwarfs with substellar companions are relatively rare. Indeed only 0.5 per cent of white dwarfs are thought to have an unresolved substellar companion \citep{steele11, Rebassa2019}. Despite extensive searches \citep{debes11, dennihy17, hogg20, Lai21, Owens23}, there are fewer than 15 white dwarfs known to have brown dwarf companions that have survived common envelope evolution, and now orbit with periods of $\sim$hrs (e.g. \citealt{littlefair2014, parsons2017, casewell2020}). A similar number of wide, often wide enough to be considered common proper motion companions are also known (e.g. \citealt{French2023, meisner, day-jones11, deacon14}). Recent results from $Gaia$ \citep{rogers24} and $JWST$ \citep{mullally24} indicate that many more companions may be found in the near future.

PHL~5038AB is a resolvable binary comprising a cool white dwarf with a L-dwarf companion separated on the sky by 1". The white dwarf was first identified by \citet{eisenstein} with \citet{steele09} discovering the brown dwarf companion, which at the time, was only the fourth known to orbit a white dwarf after GD~165AB \citep{becklin88}, WD0137-349 \citep{maxted06} and GD~1400AB \citep{farihi04}. We re-observed the PHL5038AB system 12 years after \citet{steele09} in order to begin orbital monitoring of the binary. We present here a new analysis of the system including refined parameters for both the white dwarf and the brown dwarf making use of the parallax from $Gaia$ \citep{gaia}, and new spectroscopy from XSHOOTER on the Very Large Telescope. Our analysis shows calcium pollution within the white dwarf atmosphere suggesting the brown dwarf companion is disrupting the orbits of planetesimals that are disintegrating into the white dwarf atmosphere, suggesting that this mechanism indeed is possible (e.g. \citealt{Bonsor15}).

\section{Observations and data reduction}

\subsection{Gemini/NIRI Imaging} \label{sec:niri}

We imaged PHL~5038 on 2021-05-18~UT with the Near Infrared Imager (NIRI; \citealt{2003PASP..115.1388H}) in the $K_s$ band as part of programme GN-2021A-FT-207 at the Frederick C.\ Gillett Gemini Telescope (Gemini North). We obtained 14 60~s exposures at an airmass of 1.09 with the f/6 camera providing a pixel scale of 0.117\arcsec\,pixel$^{-1}$. 

We reduced the data using the \textsc{dragons} software \citep{dragons} provided by Gemini Observatory. Flat field and dark frames were obtained for our observations as part of daytime calibrations, and we used \textsc{dragons} to create a bad pixel mask from the 10~s dark frames. The 14 images were reduced and stacked using stars in the image as references. We also stacked the images using the header coordinate system but there was negligible difference between the methods.

We re-reduced the NIRI acquisition images presented in  \citet{steele09}.  These images were taken in 2008 in the $K$ band and were reduced using \textsc{dragons} as for the newer NIRI data.

\subsection{XSHOOTER}
There are two sets of observations of PHL~5038AB taken with XSHOOTER \citep{2011A&A...536A.105V} on the European Southern Observatory's Very Large Telescope as part of programmes 384.D-0494 (PI: Steele) and 106.213V (PI: Rebassa-Mansergas).
The data from the Steele programme were taken on 19-10-2009 with 750~s exposures in the UVB and VIS arms in seeing of 0.93\arcsec\ and airmass of 1.  Data for the Rebassa-Mansergas programme were taken on 14-10-2020 with exposure times of 450~s in seeing of 0.68\arcsec\ and airmass of 1.1. All observations used the 0.9\arcsec\ slit and the nodding mode.

We used version 2.11.5 of the ESOReflex \citep{2013A&A...559A..96F}\footnote{\url{http://www.eso.org/sci/software/esoreflex/}} data reduction workflow, implementing the XSHOOTER pipeline version 3.5.3, to reduce, combine and calibrate the spectra. The 2010 spectra were combined to create a single 1500~s exposure with S/N of $\sim$50, and the 2020 spectra were combined to create a 900~s exposure with S/N of $\sim$20.

\section{Results}
\subsection{White dwarf parameters}
\citet{steele09} determined T$_{\rm eff}$=8000$\pm$100~K and $\log{g}$=8.2$\pm$0.2 from the Sloan Digital Sky Survey (SDSS) spectrum \citep{sdss} implying a mass of 0.72$\pm$~0.15M$_{\odot}$ and distance of 64$\pm$10 pc from the flux scaling factor for the white dwarf. However, as the effective temperature is below 12,000~K, these values are now known to be an overestimate due to the lack of 3D corrections to white dwarf models available at the time \citep{tremblay11}.

\citet{Anguiano} used the SDSS DR12 spectra to determine T$_{\rm eff}=$7575$\pm$57~K, $\log{g}$ = 7.59$\pm$0.13 with a mass of 0.41$\pm$0.07~M$_{\odot}$ %This mass is considerably smaller than that of \citet{steele09}, and these data also led \citet{Anguiano} 
and determined the photometric distance to PHL5038AB to be 96$\pm$10~pc. The $Gaia$ DR3 distance is measured to be 73.5$\pm$0.6~pc, notably different to both the \citet{steele09} and the \citet{Anguiano} distance. The parameters from \citet{gentile} based on the spectral energy distribution and the $Gaia$ eDR3 distance are T$_{\rm eff}$=8021$\pm$186~K, log g=8.03$\pm$0.07 and $M$=0.61$\pm$0.04~M$_{\odot}$. A recent photometric analysis was also performed by \citet{raddi22} who used $Gaia$ broad band photometry combined with the $Gaia$ eDR3 parallax to obtain 
T$_{\rm eff}$=7762$\pm$100~K, $\log{g}$ = 7.95$\pm$0.04 with a mass of 0.56$\pm$0.02~M$_{\odot}$.  A spectral fit was performed by \citet{kilic20} and gave T$_{\rm eff}$=7525$\pm$25~K, $\log{g}$ = 7.89$\pm$0.02 with a mass of 0.53$\pm$~0.02M$_{\odot}$, values consistent with the work of \citet{tremblay11}.  More recently, using the $Gaia$ DR3 spectrum, \citet{Jimenez2023} determined an effective temperature of 7735$\pm$150 K, a $\log{g}$=7.94$\pm$0.05 and $M$=0.56$\pm$0.03 M$_{\odot}$, in agreement with the values of \citet{gentile} and \citet{kilic20}.
  %This work did, however determine a radial velocity for the white dwarf of 44.33$\pm$5.45~Kms$^{-1}$. 

We fit the optical spectrum from XSHOOTER following the method in \citet{Rebassa2007} and determine T$_{\rm eff}$ = 7800$\pm$50~K, $\log{g}$ = 7.97$\pm$0.03 dex, $M$ = 0.574$\pm$0.018 M$_{\odot}$ and $R$=0.0129$\pm$0.0002 R$_{\odot}$ using the models of \citet{Koester2010} and the 3D correction of \citet{tremblay11}. Our results are consistent with those of  \citet{kilic20} and \citet{Jimenez2023}.  We adopt the \citet{kilic20} parameters in this work as our abundance models use the same model grid, and our fit to the XSHOOTER spectrum with the \citet{Koester2010} models gives a consistent result. 

The white dwarf parameters and the distance measurement have changed since the work of \citet{steele09}, due to higher resolution spectra with a better signal-to-noise ratio, and increasingly detailed atmospheric models including 3D corrections. Most notably the measurement of the surface gravity of the white dwarf has been revised to a lower value, now that 3D corrections have been applied, and the derived mass has decreased due to the lower $Gaia$ distance, meaning the age estimate for the system and hence the progenitor mass have changed considerably. 

We used the \textsc{wdwarfdate} software (\citealt{kiman22} \footnote{\url{https://github.com/rkiman/wdwarfdate}}) to determine the cooling age and likely mass of the white dwarf progenitor as well as the total system age. We used our white dwarf parameters with a DA white dwarf model and the \citet{cummings} initial mass-final mass relation derived using the MIST isochrones \citep{Choi2016} as input to the code. The resultant fits gave a cooling age of 1.14$^{+0.05}_{-0.04}$~Gyr, and a total system age of 10.26$^{+3.09}_{-3.61}$~Gyr with a white dwarf progenitor mass of 1.07$^{+0.17}_{-0.08}$~M$_{\odot}$. The errors on the progenitor mass and lifetime are dominated by the fact that semi-empirical initial mass-final mass relations for white dwarfs heavily rely on open cluster white dwarfs, which tend to have masses that are larger than average, due to their relative youth compared to field white dwarfs. The \citet{karakas} models for Asymptotic Giant Branch (AGB) stars at first thermal pulse give an initial mass of 1.0 M$_{\odot}$ for a core mass of 0.53 M$_{\odot}$ which is consistent with the \textsc{wdwarfdate} results for all metallicities.

\begin{table}
\centering
\caption[]{Parameters for PHL~5038AB used in this paper.} \label{tbl:data}
\renewcommand{\arraystretch}{1.5}
\begin{tabular}{lcc}
\hline
Parameter & PHL~5038A & PHL~5038B\\
\hline
Distance (pc)& 75.5$\pm$0.6& 75.5$\pm$0.6\\
T$_{\rm eff}$(K)&7525$\pm$25& 1425\\
log g&7.89$\pm$0.02&5.454\\
Mass (M$_{\odot}$&0.53$\pm$0.02&0.07\\
T$_{\rm cool}$ (Gyr)&1.14$^{+0.05}_{-0.04}$& - \\
Age (Gyr)&10.26$^{+3.09}_{-3.61}$&10.26$^{+3.09}_{-3.61}$\\
M$_{\rm {init}}$ (M$_{\odot}$)& 1.07$^{+0.17}_{-0.08}$&-\\
Accretion rate (Ca:gyr$^{-1}$)& 1.32$\times$10$^{13}$ &-\\
Accretion rate (Chondritic:gs$^{-1}$) & 7.4$\times$10$^{6}$&-\\
Accretion rate (Bulk Earth:gs$^{-1}$) & 2.6$\times$10$^{7}$&-\\
\hline
\end{tabular}
\end{table}

\subsection{Brown dwarf parameters} 
Our new, older age estimate for the PHL~5038AB system means that the 60 M$_{\rm{Jup}}$ mass determination for the brown dwarf from \citet{steele09}is also likely an underestimate. \citet{steele09} determined a spectral type of L8-L9 using a $HK$ spectrum from NIRI on Gemini North. Using the effective temperature vs spectral type relations in \citet{dupuy17} we determine the effective temperature of spectral type L8-9 to be 1300-1450~K. The \citet{Marley21} models at an age of 8~Gyr predict masses of 0.068-0.070~M$_{\odot}$ for PHL~5038B.

Using the bolometric luminosity derived from the $K$ magnitude for the brown dwarf in \citet{steele09} and the relations in \citet{dupuy17} we compared to the Sonora Bobcat models \citep{Marley21} determining T$_{\rm eff}$=1577~K and a mass of 0.071~M$_{\odot}$ for an age of $\sim$8~Gyr. An older age of $\sim$10.5~Gyr gives an effective temperature of 1425~K and a mass of 0.070~M$_{\odot}$ which is consistent with the effective temperature suggested in \citet{dupuy17} for the spectral type.

\subsection{Astrometry and orbital parameters}

 We measured relative astrometry for PHL~5038AB using the stacked images for each epoch. We fitted an analytic point spread function model to each component, with the model using three concentric 2D Gaussians with different amplitudes, standard deviations, ellipticities, and angles for the ellipticities. This approach is based on work with adaptive optics imaging of low-mass binaries (e.g., \citealt{Liu06}). We then converted the positions in pixels of the two binary components into the sky coordinates using the WCS information provided by the telescope in the FITS headers. The system is resolved in both epochs, and the errors in the relative astrometry are dominated by the astrometric calibration of NIRI. We adopt the same uncertainties as in \citet{mann19} for their large sample of NIRI astrometry: a fractional uncertainty of 0.23 per cent in pixel scale and 0.1$^{\circ}$ in the detector orientation. 

\begin{figure*}
	\includegraphics[scale=1, trim={0 2cm 1cm 3cm},clip] {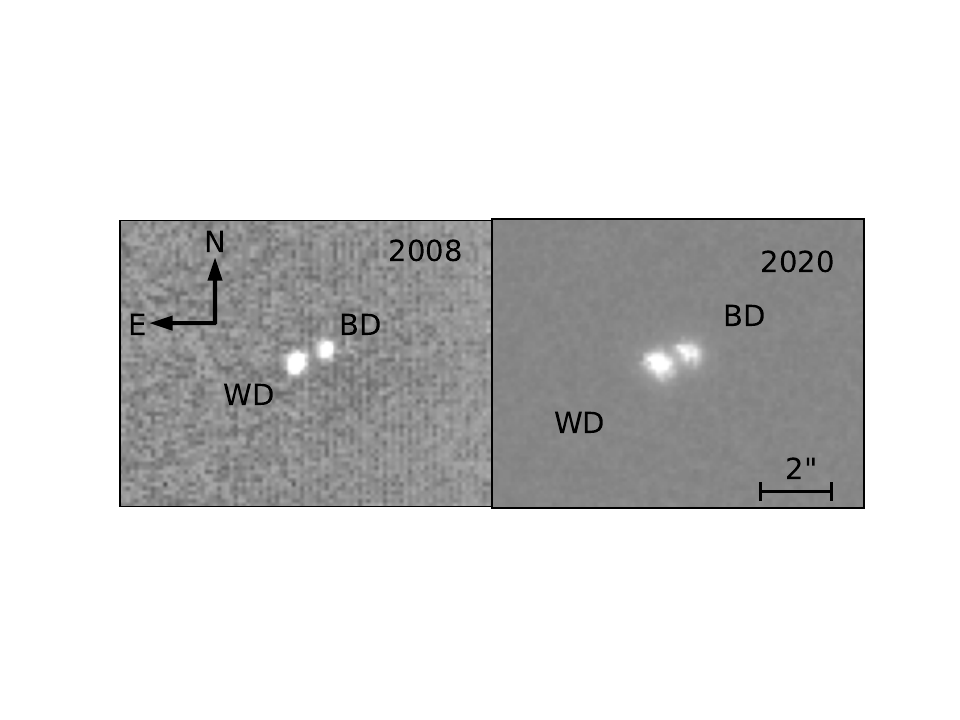}
    \caption{NIRI images from 2008 (left) and 2020 (right) showing the positions of the white dwarf and the brown dwarf at each epoch.}
    \label{image}
\end{figure*}

We determined the contrast between the two sources is $0.63 \pm 0.03$\,mag (Table \ref{tbl:relast}). Using the white dwarf magnitudes predicted by \citet{holberg} and \citet{bergeron95} for a 7500~K white dwarf with $\log{g}=8.00$\,dex (cgs), we obtain a $K_s$ absolute magnitude of 12.48. While no errors are given on this synthetic photometry, the difference in magnitude between a 7000~K and a 7500~K white dwarf is 0.08. The brown dwarf absolute magnitude for an L8 dwarf from fitting photometry for known brown dwarfs with accurate parallaxes \citet{dupuy12} is $K_s=13.06\pm0.03$\,mag.  While our measured difference in photometry is only $\sim$0.6 mags, the combined errors on the synthetic photometry are less than 0.1 mags, meaning the white dwarf is the brighter component to the southeast in Figure \ref{image}.

\begin{table}
\centering
\caption[]{Gemini/NIRI relative astrometry and photometry of PHL~5038AB.} \label{tbl:relast}
\begin{tabular}{lcccc}
\hline
Epoch & Separation (mas) & Position Angle ($^{\circ}$) & $\Delta{m}$ (mag)& Band \\
\hline
2008.6132 & $908\pm5$ & $294.66\pm0.1$ & $0.63\pm0.03$ &$K_s$\\
2021.3774 & $887\pm5$ & $290.48\pm0.1$ & $0.63\pm0.03$ & $K$\\
\hline
\end{tabular}
\end{table}

We used the Markov-Chain Monte Carlo (MCMC) orbital analysis tool {\sc orvara} \citep[v1.0.4;][]{2021AJ....162..186B} to fit orbits to our relative astrometry for PHL~5038AB. We only used one informative prior on the total mass of the system, based on the white dwarf mass estimate of $0.53\pm0.03$~M$_{\odot}$ and companion mass estimate of $0.070\pm0.002$~M$_{\odot}$. This version of {\sc orvara} does not allow total mass priors, but when fitting relative astrometry that only constrains the total mass, setting a prior on the primary mass of $0.60\pm0.03$~M$_{\odot}$, and a zero mass prior on the companion, is functionally equivalent to employing a total mass prior. For the remaining orbital elements, we used their default priors: linear-flat in the eccentricity ($e$) and viewing angles (except inclination, $p(i)\propto\sin{i}$), and log-flat in semimajor axis $a$. Our results are based on a run with 100 walkers, $10^6$ steps for the MCMC, and 5 temperatures for parallel tempering. We thinned our chains, retaining every 50th step, and discarded the first 75 per cent as burn-in, yielding $5\times10^5$ final samples in our posterior.

\begin{figure}
	\includegraphics[width=\columnwidth]{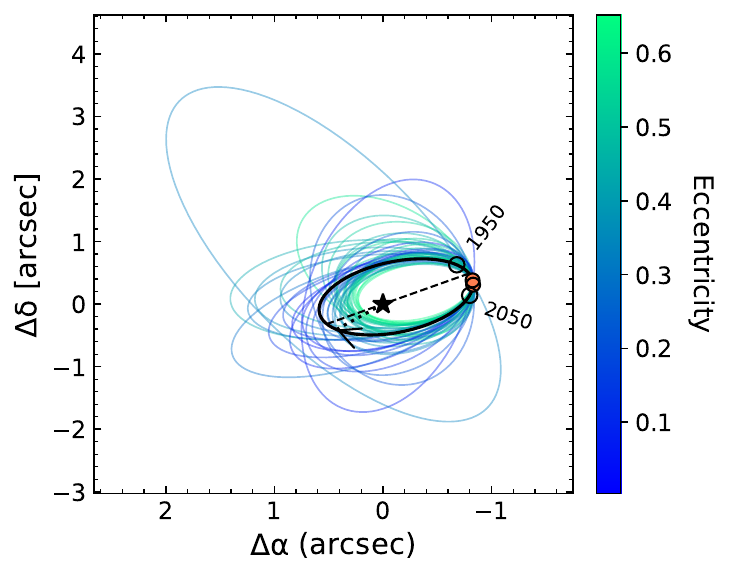}
 \vskip -0.1in
    \caption{Randomly drawn relative orbits for PHL~5038B drawn from the MCMC posterior with the highest-likelihood orbit overplotted (thick black line). The white dwarf PHL~5038A is represented by the black star and PHL~5038B by the orange dot showing the movement between the two epochs of data.}
    \label{fig:orbit}
\end{figure}

The posterior distributions of orbital parameters correspond to semimajor axes of $66^{+12}_{-24}$~AU, and inclinations of $132\pm11^{\circ}$. The eccentricity is essentially unconstrained due to the large errors on the inclination, except that the posterior drops off steeply around the 2-$\sigma$ upper limit of $e<0.615$  (Figure \ref{fig:orbit}). 

\subsection{PHL~5038A: A polluted white dwarf}

\begin{figure*}
	\includegraphics[scale=1, trim={0cm 13cm 6cm 2.27cm},clip] {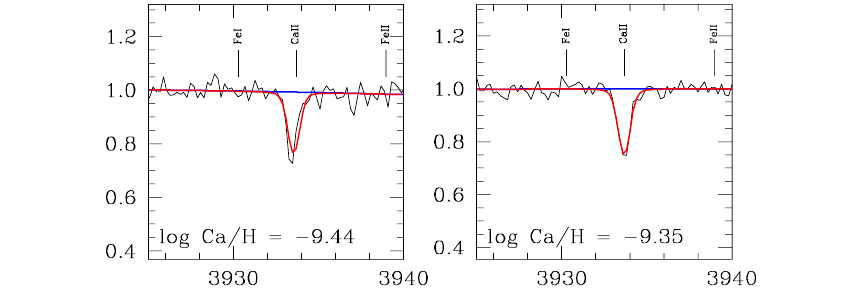}
    \caption{The Ca K line fit with a DA white dwarf model using the parameters of \citet{kilic20}
    following the methods of \citet{dufour07} and \citet{coutu19}. The [Ca/H] abundances are given for each epoch.}
    \label{fig:ca}
\end{figure*}

Using the white dwarf parameters from \citet{kilic20}, we fit both epochs of spectra using the 3934\,\AA\ Ca~\textsc{ii} K line and the models of \citet{dufour07} and \citet{coutu19} to determine the [Ca/H] abundance (Figure \ref{fig:ca}). The abundances are -9.44 for the 2008 spectrum and -9.35 for the 2020 spectrum.  We also used \textsc{molly} \footnote{\url{https://cygnus.astro.warwick.ac.uk/phsaap/software/molly/html/INDEX.html}} to measure the equivalent width of the line, measuring a value of 0.246$\pm$0.009  and 0.231$\pm$0.024 \AA\, for the two epochs, determining there is no significant variability in abundance between the two epochs.

We also determined the radial velocity of each epoch using the Ca~\textsc{ii} K absorption line and obtained measurements of 70.46 and 69.07\,km\,s$^{-1}$, which results in radial velocities of  23.46 and 23.69\,km\,s$^{-1}$ once the heliocentric corrections (-22.39 and -20.82\,km\,s$^{-1}$) and gravitational redshift (24.6$\pm$1\,km\,s$^{-1}$) are applied.  
These radial velocities are consistent with the 17.59$\pm$14.21\,km\,s$^{-1}$ determined by \citet{raddi22}, and the 44.33$\pm$5.45\,km\,s$^{-1}$ from \citet{Anguiano} if it is corrected by our gravitational redshift value. These results are also shown in Table \ref{tbl:abun}.

This abundance is consistent with other cool DAZ white dwarfs in \citet{Zuckerman03} - LHS3007 (T$_{\rm eff}$=7366~K; log g = 7.58) has a similar [Ca/H]=-9.312, as does the more massive GD96 (T$_{\rm eff}$=7373~K, log g = 8.00), [Ca/H]=-9.409. The abundance is also consistent with the effective temperature [Ca/H] relation in \citet{Blouin}, indicating that while this abundance is lower than that of many of the well-known polluted white dwarfs hosting discs, it is not particularly unusual for a white dwarf with PHL~5038A's parameters.

\subsection{Accretion rates}
We used the Montreal White Dwarf Database \footnote{\url{https://www.montrealwhitedwarfdatabase.org}} \citep{dufour17} and the \citet{kilic20} values of effective temperature and surface gravity and our average measured [Ca/H] of -9.36 to determine the convection zone mass ratio (log CVZM=-8.298) and the diffusion timescale of calcium in the white dwarf atmosphere (log Ca$_{\rm settle}$ =3.848). We then follow the method in \citet{Koester09}. 

\begin{table}
\centering
\caption[]{Abundance measurements for PHL~5038A.} \label{tbl:abun}
\begin{tabular}{lcc}
\hline
Measurement & 2008 spectrum& 2020 spectrum\\
\hline
[Ca/H]&-9.44& -9.35\\
EW(\AA)&0.246$\pm$0.009&0.231$\pm$0.024\\
RV (kms$({-1}$)&23.46&23.69\\
\hline
\end{tabular}
\end{table}

For PHL~5038A, the accretion rate of calcium is  1.32 $\times 10^{13}$ g\,yr$^{-1}$, which if we assume is due to accretion of bodies with chondritic abundance (calcium fraction of 0.057), equates to a total accretion rate of  $2.32 \times 10^{14}$ g\,yr$^{-1}$, or $7.4 \times 10^{6}$ g\,s$^{-1}$ (Table \ref{tbl:data}). If we compare this accretion rate to the total amount of Ca expected within the convection zone of the white dwarf, we determine a value of $\sim$10$^{18}$~g.  Similarly, this accretion rate is $2.6 \times 10^{7}$ g\,s$^{-1}$ if we assume the abundance of bulk Earth with 1.6 per cent calcium abundance.  

\begin{figure}
	\includegraphics[scale=0.4]{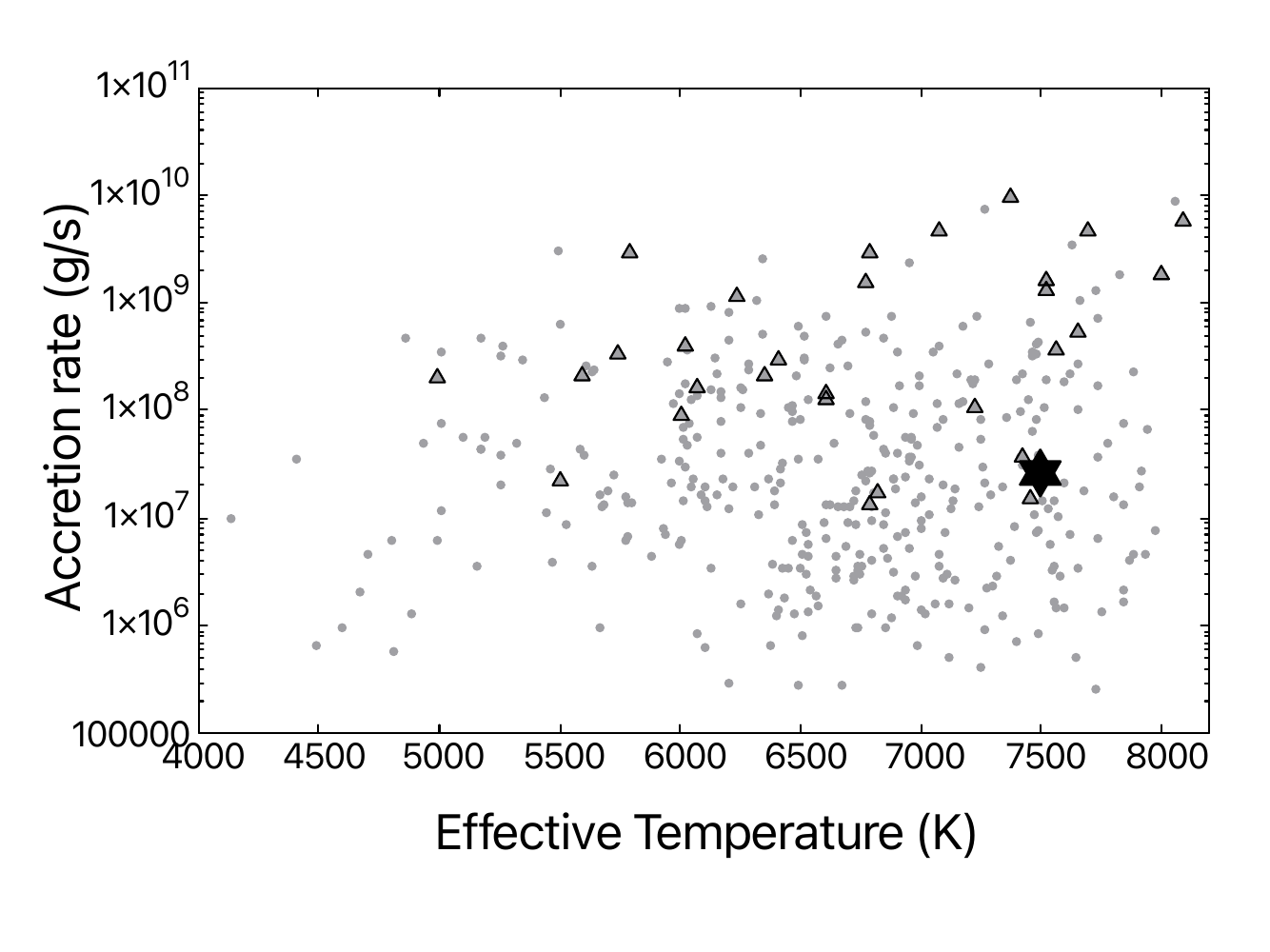}
    \caption{Effective temperature vs accretion rate for polluted white dwarfs from \citet{Blouin}. DZ (helium atmosphere) white dwarfs are plotted as grey points, DAZ (hydrogen atmosphere) white dwarfs as outlined triangles. PHL~5038A is plotted as the black square.}
    \label{fig:pollution}
\end{figure}

This accretion rate is low, but comparable to other DZ white dwarfs with similar cooling times (Figure 4 in \citealt{Blouin}): grey dots in Figure \ref{fig:pollution}), but there are very few cool DAZ white dwarfs with similar accretion rates ($\sim$30 triangles on Figure \ref{fig:pollution}), with PHL~5038A having a lower accretion rate than the majority.

\subsection{A debris disc?}
To determine whether there was an infrared excess indicative of a debris disc, we performed an SED fit by using the SDSS photometry apart from the $u'$ band, which can suffer from both reddening and atmospheric issues, and the resolved near-infrared photometry from \citet{steele09}. To determine the white dwarf parameters we %followed the method in \citet{steele21} and 
performed a chi-square minimisation on an interpolated grid of white dwarf cooling models \footnote{\url{https://www.astro.umontreal.ca/~bergeron/CoolingModels/}} from \citet{bedard} and \citet{blouin18}.

\begin{figure*}
	\includegraphics[scale=0.4]{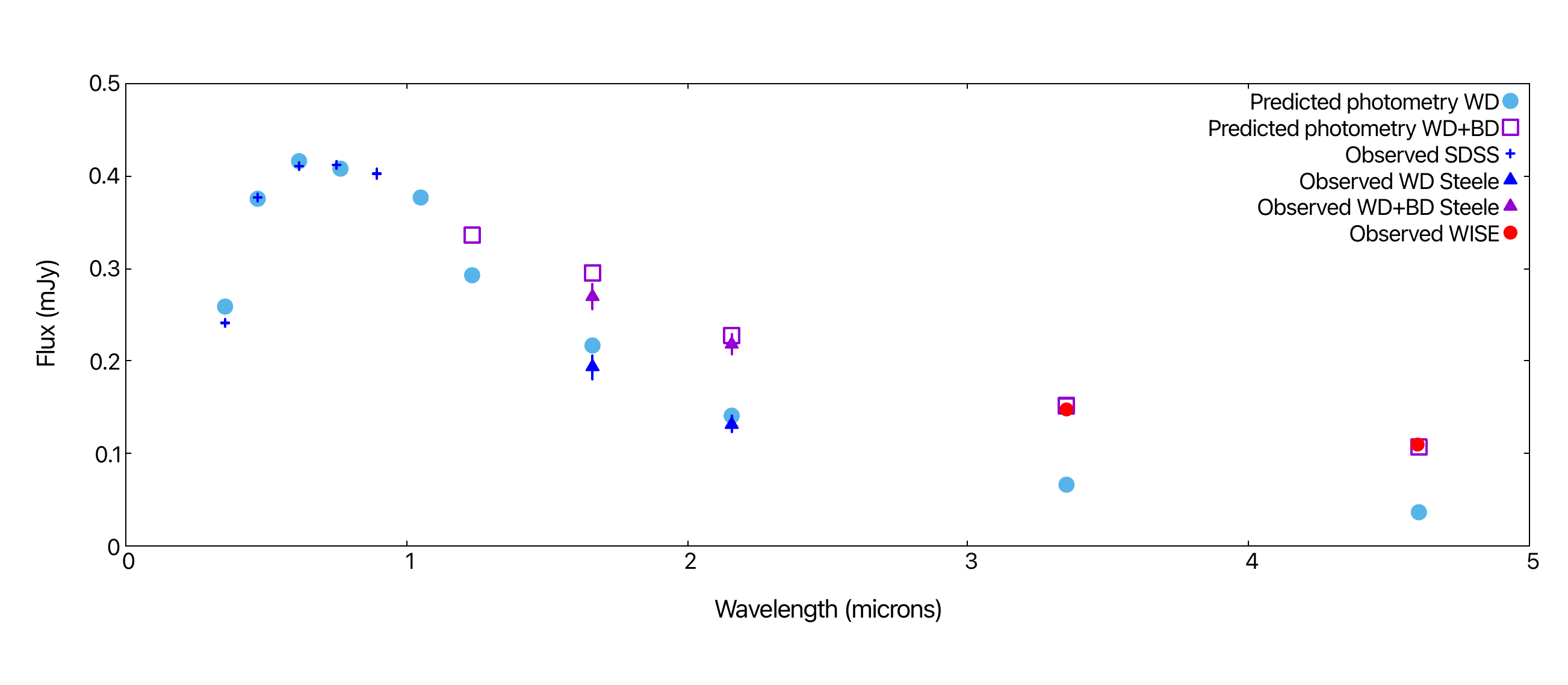}
    \caption{Observed and predicted photometry for PHL~5038AB. The observed SDSS photometry for the white dwarf is shown as dark blue pluses, and the observed near-IR photometry from \citet{steele09} is shown as filled triangles for the white dwarf (dark blue) and the combined white dwarf-brown dwarf (purple). The observed WISE magnitudes for the combined white dwarf--brown dwarf are shown as filled red circles. Our predicted photometry for the white dwarf is shown as filled light blue circles. The predicted photometry for the combined system is shown as purple boxes. There is no infrared excess that could be due to a disc detected out to 4.5 microns.}
    \label{fig:excess}
\end{figure*}

We determined that PHL~5038A should have T$_{\rm{eff}}$=7751~K with a 95 per cent confidence interval between 7460 and 8045~K, and log g of 7.94 (95 per cent confidence interval 7.84 to 8.04), and a mass of 0.56 M$_{\odot}$ (95 per cent confidence interval: 0.51 to 0.62~M$_{\odot}$), which is consistent with the \citet{kilic20} spectroscopic fit. We also used the synthetic photometry for DA white dwarfs in the NIR/WISE bands from \citet{holberg} to determine the SED of the white dwarf. To determine the likely magnitudes of the brown dwarf companion we used the observed absolute magnitudes of brown dwarfs from \citet{dupuy12} scaled to the observed $K$ band flux of the brown dwarf to take into account the large rms scatter in the relationship for L8-L9 dwarfs (L8: 0.20\,mag, L9: 0.43\,mag).
Figure \ref{fig:excess} shows that the detected excess in the $JHK$ and $Wise$ magnitudes is due to the brown dwarf companion alone.

In order to put limits on whether an undetected disc could be present in the system we used the estimated uncertainty in the brown dwarf fluxes from the SED calculation  
added in quadrature to the measured flux uncertainty from the photometry to adopt a 3-$\sigma$ excess criterion above the W1 and W2 observed fluxes. We then estimated the limits on any disc present. We initially considered optically thick discs using the models from \citet{Jura2003}. The Roche limit for PHL~5038A is $\sim$48 R$_{\rm WD}$, and the sublimation temperature of $\sim$1600~K is reached at $\sim$5~R$_{\rm WD}$. Assuming these radii are the inner and outer radii for an optically thick disc, all such discs are ruled out to nearly edge-on, $<$89.9 degrees inclination. If we instead consider very narrow rings that are near face-on, a 1~R$_{\rm WD}$ WD thick ring at $<$33 R$_{\rm WD}$ is ruled out. Narrow rings at smaller radii will be allowed for increasing inclination limits, so are not completely ruled out but are likely to be rare.  We therefore conclude that an optically thick disc is unlikely to be present.

We then considered optically thin discs. We used black body discs with temperatures between 600 and 1200~K, spanning the most typical white dwarf disc temperatures in the temperature regime where the W1 and W2 wavelengths are most sensitive. We then calculated the effective limiting area for such discs to be detectable at our 3-$\sigma$ excess criterion. We find that for 600~K the limiting emitting area is 0.70\,R$_{\odot}^2$ and for 1200~K, it is 0.17\,R$_{\odot}^2$. If we assume a single grain radius for the emitting dust and a rocky composition such that $\rho _{\rm dust}$=3\,g\,cm$^{-3}$, then these areas correspond to  9.4 $\times$ 10$^{17}$ (R$_{\rm dust}$) g at 600~K and 5.9 $\times$ 10$^{16}$ (R$_{\rm dust}$) g at 1200 K where R$_{\rm dust}$ is the radius of the dust particles.

\section{Discussion}

The lack of detection of a disc around PHL~5038A is perhaps not unexpected. For instance, while 25-40 per cent of white dwarfs show signs of metal pollution, only 1.5-4 per cent of white dwarfs show an excess due to a debris disc. It is unlikely the pollution is caused by wind accretion from the brown dwarf.
None of the white dwarf primaries in the close white dwarf-brown dwarf binaries with orbital periods of $\sim$1 hrs shows any metal pollution in their atmospheres, meaning this mechanism is unlikely to have a significant effect on a binary with separation of $\sim$70 AU.

Figure 9 of  \citet{bonsor17} suggests our accretion rate for Ca, combined with the low effective temperature of the white dwarf, means PHL~5038A falls close to the boundary between their region A, where the dust has been totally accreted, and region B, where the dust is optically thin and is dominated by Poynting-Robertson drag. \citet{bonsor17} suggest that for white dwarfs with a sinking timescale of longer than $\sim$500~yrs (of which PHL~5038A is one) then a finite dust lifetime can provide an explanation for a lack of detectable disc.  

\citet{hollands18} suggest for DZ white dwarfs where no disc is detected the accretion phase may have indeed finished,  and the detected metals are slow sinking tracers of a previous accretion event. However, while PHL~5038A is relatively cool, it is not a DZ. The calcium diffusion timescale is $\sim$7000~yrs, and the  average lifetime of a disc around a white dwarf is predicted to be between 3$\times10^4$ and 5$\times10^6$ years assuming accretion is at a constant rate \citep{girven}. These values would suggest that the accretion phase has not yet ended. Indeed, the level of accretion in this cool, polluted white dwarf is sufficiently low that it could also be supplied by an optically thin dust disc accreting via Poynting-Robertson drag without the dust disc producing a detectable infrared excess, or via a mechanism such as that suggested by \citep{Metzger2012}. 

\citet{brouwers22} however, showed that accretion via Poynting-Robertson drag is unable to produce the necessary accretion on short enough timescales to replicate the abundances detected for many white dwarfs. They place the limit at 10$^{6}$\,g\,s$^{-1}$, stating that in order to achieve an accretion rate higher than this, the larger asteroids must be ground down into an eccentric tidal dust disc, perhaps via perturbations caused by a giant planet. Our accretion rate is larger, at 7.4$\times10^{6}$\,g\,s$^{-1}$, but still low for a white dwarf. \citet{brouwers22} also predict that lower accretion rates are more likely to occur from smaller asteroids, that take longer timescales to grind down and accrete onto the white dwarf. Assuming this is the case, \citet{kenyon17} found that if the material lost by accretion onto the white dwarf is continually replenished, an equilibrium mass might be achieved for the disc. Assuming a collisional cascade at the tidal disruption radius of the white dwarf, the output of gas accretion onto the white dwarf is then equivalent to the influx of mass into the disk in a steady state. If we use the default values in equation 11 of \citet{kenyon17}, with our calculated accretion rate for PHL~5038A, we find the equilibrium mass is $=3.6\times10^{17}$~g assuming particles of radius 1~km. Interestingly, this value also corresponds to just above the accretion rate predicted to show an infrared excess assuming late-stage dust accretion from highly eccentric asteroids \citep{brouwers22}.
As we have no detectable disc, we cannot put limits on the grain sizes present, but we can combine the estimate from the collisional cascade, with the maximum emitting area we derived in Section 3.6, where we assumed a single grain radius. This combination gives a lower limit of  9.4 $\times 10^{11}$ g at 600~K, assuming the particle size is 1~micron. The upper limit is found by multiplying this mass by $\sqrt{1~\rm km/1~\rm micron }$, to 3$\times10^{16}$ g.

The arguments presented in \citet{burleigh02} suggest that PHL~5038B has not always orbited the white dwarf at 66$^{+12}_{-24}$~AU, and has likely increased its orbital separation as the white dwarf progenitor moved off the main sequence 1.14$^{+0.05}_{-0.04}$~Gyr ago. \citet{Jeans24} calculates that the orbits of planets that do not interact directly with the white dwarf progenitor as it becomes a giant (e.g. those that escape a phase of common envelope evolution) will simply expand their orbits adiabatically by a maximum factor of $\rm {M_{MS}/M_{WD}}$. For PHL~5038AB this suggests that the orbit expanded by a factor of two, placing the initial orbit at $\sim$33 AU, not dissimilar to the location of Neptune in our own solar system.   The models of \citet{ventura} suggest that the maximum radius of an AGB star of mass 2.5~M$_{\odot}$ (the lowest mass they present) should be $\sim$2.5~AU, considerably smaller than the predicted initial orbit of the brown dwarf confirming this system has had no common envelope phase.

Of the 11 post-common envelope binaries comprising a white dwarf and a brown dwarf, and the $\sim$10 wider systems (e.g., \citealt{meisner} and references therein), only one other white dwarf shows any sort of metal pollution, SDSS~J155720.77+091624.6A \citep{farihi17}. This white dwarf has a brown dwarf companion on an orbital period of 2.73~hr and the distinctive low mass of the white dwarf is indicative of a post-common envelope system.  This binary has a mid-infrared excess seen at 3.6 and 4.5 microns that cannot be attributed to a reflection effect or the white dwarf \citep{swan20, farihi17} and is determined to be emission from a debris disc.  However, due to the short period of the system, stable circumstellar material is only permitted at a radius very close to the white dwarf, where the dust would be inconsistent with the observed thermal emission, thus indicating the disc is circumbinary, and is located at $\sim$3.3~R$_{\odot}$ or 0.015~AU.

Unlike SDSS~J1557, PHL~5038AB is widely separated, meaning it is possible for debris to be present on a stable S-type orbit between the white dwarf and the brown dwarf. Using Equation~1 of \citet{holman99} for our orbital parameters, we took the zero eccentricity case, determining any debris would be stable at a distance  $<$17--32\,AU. If we take the maximum eccentricity of 0.6 suggested by the orbital fit, the debris is stable much closer to the white dwarf, closer than 5--8\,AU, a significantly narrower range.  These values for possible locations of any debris are all outside the typical radius of an AGB star, suggesting the debris may have been a remnant from the binary's formation.

It is therefore possible that the presence of the brown dwarf is responsible for the pollution seen in the white dwarf atmosphere. If the debris belt is currently at the ``edge'' of the stable zone, with the outer edge of the belt being slowly eroded by interactions with the brown dwarf, the belt could have initially been larger. In this scenario, when the white dwarf was first formed, there would have been an intense period of scattering as the outer edge of the belt was cleared. This scattering decreased with time, leaving us with the low calcium abundance detected in the white dwarf atmosphere.

\section{Conclusions}
We determine that the white dwarf PHL~5038A is polluted by calcium, possibly by rocky material that is being perturbed by the wide brown dwarf companion PHL~5038B which orbits at $66^{+12}_{-24}$~AU. The brown dwarf likely orbited the $\sim$1~M$_{\odot}$ white dwarf progenitor at $\sim$33~AU, and migrated outwards as the star evolved off the main sequence, without a common envelope phase. PHL~5038AB is perhaps the first system with a wide substellar companion that could be responsible for the pollution seen at the white dwarf.

\section*{Acknowledgements}
We thank Kathleen Labrie for her expert knowledge of DRAGONS and assistance with the data reduction. We also thank (and greatly miss) Tom Marsh for the use of Molly.

Based on observations collected at the European Organisation for Astronomical Research in the Southern Hemisphere under ESO programme(s) 106.213V.001 and 0106.D-0386(A). Based on observations obtained at the international Gemini Observatory, a program of NSF's NOIRLab, which is managed by the Association of Universities for Research in Astronomy (AURA) under a cooperative agreement with the National Science Foundation on behalf of the Gemini Observatory partnership: the National Science Foundation (United States), National Research Council (Canada), Agencia Nacional de Investigaci\'{o}n y Desarrollo (Chile), Ministerio de Ciencia, Tecnolog\'{i}a e Innovaci\'{o}n (Argentina), Minist\'{e}rio da Ci\^{e}ncia, Tecnologia, Inova\c{c}\~{o}es e Comunica\c{c}\~{o}es (Brazil), and Korea Astronomy and Space Science Institute (Republic of Korea). The data have been processed using \textsc{dragons} (Data Reduction for Astronomy from Gemini Observatory North/South).This research has also made use of the Spanish Virtual Observatory (\url{https://svo.cab.inta-csic.es}) project funded by MCIN/AEI/10.13039/501100011033 through grant PID2020-112949GB-I00.

ARM acknowledges support from the AGAUR/Generalitat de Catalunya grant SGR-386/2021 and from the Spanish MINECO grant PID2020-117252GB-I00. RMO is funded by INTA through grant PRE-OBSERVATORIO. JRF acknowledges the support of a University of Leicester College of Science and Engineering Studentship.  RDA acknowledges funding from the Science \& Technology Facilities Council (STFC) through Consolidated Grant ST/W000857/1. SLC acknowledges support from an STFC Ernest Rutherford Fellowship ST/R003726/1.

%%%%%%%%%%%%%%%%%%%%%%%%%%%%%%%%%%%%%%%%%%%%%%%%%%
\section*{Data Availability}

All data presented in this paper are available in public data archives (e.g., ESO, Gemini), and we use no proprietary code.



\bibliographystyle{mnras}
\bibliography{example} % if your bibtex file is called example.bib

%\begin{thebibliography}{99}
%\end{thebibliography}

%%%%%%%%%%%%%%%%%%%%%%%%%%%%%%%%%%%%%%%%%%%%%%%%%%

%%%%%%%%%%%%%%%%% APPENDICES %%%%%%%%%%%%%%%%%%%%%

%\appendix

%\section{Corner plot}
%\begin{figure}
%	\includegraphics[width=\columnwidth]{Corner_PHL.pdf}
 %   \caption{Corner plot for the best fit orbital solution.}
 %   \label{fig:orbit}
%\end{figure}

%\begin{figure}
%	\includegraphics[width=\columnwidth]{teff_7525.0_logg_7.89_feh_p0.00_vvcrit_0.0_DA_Cummings_2018_MIST_gridplot.png}
 %   \caption{Corner plot from \textsc{wddate} .}
 %   \label{fig:orbit}
%\end{figure}

%%%%%%%%%%%%%%%%%%%%%%%%%%%%%%%%%%%%%%%%%%%%%%%%%%

% Don't change these lines
\bsp	% typesetting comment
\label{lastpage}
\end{document}